# Investigation of Lightning Effects on Solar Power Plants Connected to Transmission Networks

S.Grebovic, A. Aksamovic, B. Filipović – Grčić, S. Konjicija

*Abstract*-- The increasing integration of solar power plants into transmission grids has raised concerns about their vulnerability to disturbances, particularly lightning strokes. Solar energy, while offering significant environmental and economic benefits, faces challenges when connected to transmission lines that are prone to lightning discharges. This paper investigates the impact of lightning events on solar power plants, focusing on overvoltage effects. Lightning stroke simulations were conducted at various distances from the solar power plant along the transmission line, considering scenarios with and without surge arrester. Key lightning parameters such as peak current, front time, and tail time were varied to simulate different lightning strokes. The study also includes a Fourier transform analysis of the resulting overvoltages with and without a surge arrester, along with the Hilbert marginal spectrum of these overvoltages. The results provide insights into the effectiveness of surge arresters in mitigating lightning overvoltages and highlight the importance of proper protective measures for enhancing the reliability and safety of solar power plants connected to transmission networks.

## I. Introduction

THE growing global demand for renewable energy has led to a rapid increase in the installation of solar power plants, ranging from small-scale systems to large-scale facilities. Solar energy has become one of the most prominent sources of clean and sustainable energy due to its environmental benefits, cost-effectiveness, and potential to reduce dependence on fossil fuels [1]. As this trend continues, an increasing number of solar power plants are being integrated into transmission grids to meet energy demand efficiently. While the integration of solar power plants into the transmission network offers some technical advantages, such as the higher capacity of transmission lines to accommodate large-scale energy production and the ability to transport electricity over long distances, it also presents certain challenges.

A substantial body of research has addressed the connection of solar power plants to electrical networks and their impact on both distribution and transmission systems. Numerous studies have also focused on the design, planning, and optimization of solar power plants, including site selection, system sizing, and technology integration. Additionally, significant attention has been given to the economic aspects of solar energy projects, grid stability challenges, the integration of energy storage systems, and the development of advanced inverters to enhance system efficiency and reliability, as well as addressing challenges like lightning-caused transients, protection measures, and the integration of advanced technologies [2-5].

However, less focus has been given to understanding how the specific conditions of the transmission grid, particularly disturbances, affect the operation of solar power plants. Transmission lines, as essential components of the grid, are highly exposed to environmental factors, including lightning discharges. Regions with favorable solar potential, characterized by a high number of sunny days, often coincide with areas prone to frequent lightning strikes. These environmental factors pose unique challenges to the reliable operation of solar power plants connected to the transmission grid. Main Contributions of this paper are:

- Investigation of Lightning-Caused Faults: The study thoroughly examines the impact of faults caused by lightning discharges that propagate through transmission lines and affect the operation of connected solar power plants.
- Shielding Failure Current Determination: Assessment of transmission line performance, including the maximum shielding failure current.
- Overvoltage Analysis: Precise modeling and evaluation of overvoltages with and without surge arresters.
- Surge Arrester Effectiveness: Comparative analysis demonstrating the role of surge arresters in mitigating lightning-caused overvoltages.

The rest of the paper is structured as follows: Section II describes the concept of connecting a solar power plant to the transmission grid, and analyses regions with a high number of sunny and thunderstorm days, along with the challenges of constructing solar power plants in such areas. Section III provides a detailed description of the EMTP model used for the analysis. Section IV outlines the test cases, presents and analyses the results, while Section V offers the concluding remarks of the paper.

---

This work was supported in part by the Federal Ministry of Education and Science, Federation of Bosnia and Herzegovina. Undertaken under grant agreement No. 05-35-2480-1/23.

S. Grebović is with University of Sarajevo, Faculty of Electrical Engineering, Sarajevo, Bosnia and Herzegovina (e-mail of corresponding author: sgrebovic@etf.unsa.ba ).
A. Akšamović is with University of Sarajevo, Faculty of Electrical Engineering, Sarajevo, Bosnia and Herzegovina (e-mail: aaksamovic@etf.unsa.ba ).
B. Filipović – Grčić is with University of Zagreb, Faculty of Electrical Engineering and Computing (email: Bozidar.Filipovic-Grcic@fer.hr )
S. Konjicija is with University of Sarajevo, Faculty of Electrical Engineering, Sarajevo, Bosnia and Herzegovina (e-mail: skonjicija@etf.unsa.ba )

## II. CONCEPT OF SOLAR POWER PLANT CONNECTION TO THE TRANSMISSION GRID AND ANALYSIS OF REGIONAL CHARACTERISTICS

Large-scale solar power plants connected to the transmission grid consist of several key components that enable efficient electricity production, conversion, and transmission. Each of these components has a specific function in generating and distributing electrical energy and connecting it to the power grid. One such structure is illustrated in Figure 1. Further details about topologies for large-scale photovoltaic power plants can be found in the literature [6].

a medium voltage, allowing efficient energy distribution at the medium level before connecting to the higher transmission voltage. Finally, to connect the solar power plant to the transmission grid, a transformer for increasing the voltage to transmission level is used. This transformer raises the medium voltage to the transmission level, typically 110 kV or higher, enabling the transmission of large amounts of electricity over long distances and ensuring the stability of the power system.

Figure 1 provides a comprehensive representation of the solar power plant, the transmission line to which it is connected, and the propagation of overvoltage waves caused by lightning discharges. When a lightning stroke strikes the

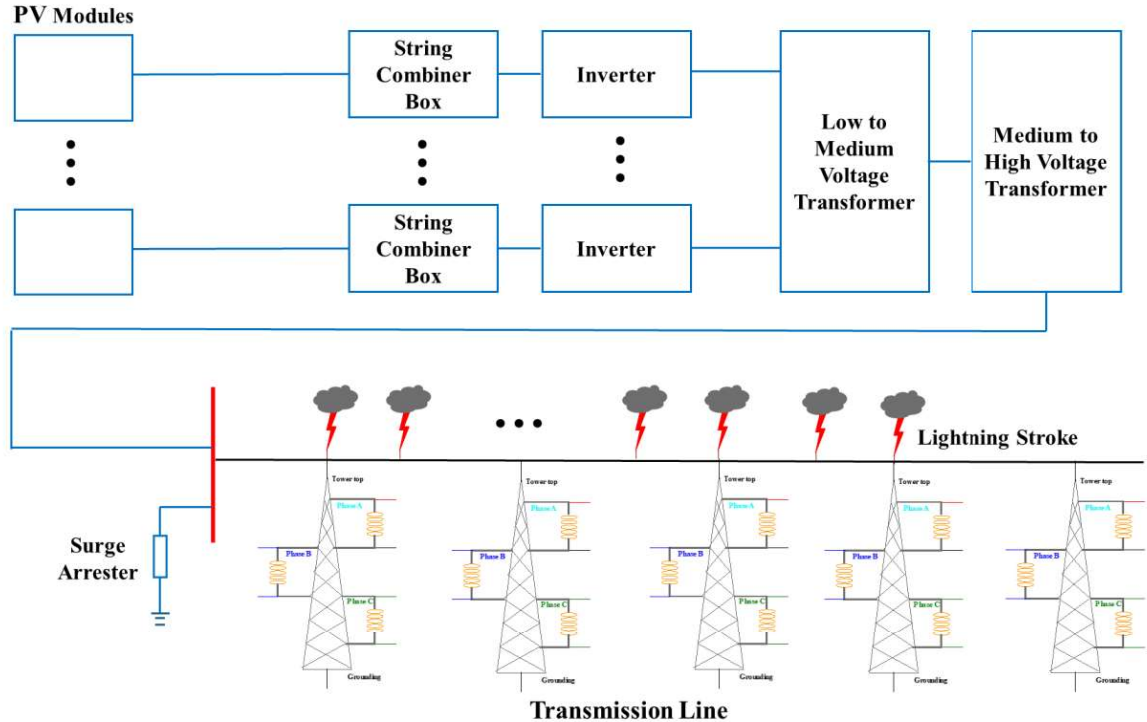

Fig. 1. Structure of the solar power plant and transmission line with illustration of the position of lightning stroke along the line

The main component of any solar power plant is the solar modules, which can be numerous and are arranged in strings. These modules directly convert solar energy into direct current (DC) electricity, forming the foundation of the entire energy production system. The current output from the modules is then collected by string combiners. These electrical devices serve to combine the output from multiple strings of PV modules and also include protective components such as fuses and surge protection, ensuring the stability and safety of the system.

Next, a key role in converting direct current (DC) into alternating current (AC) is performed by the inverter. Inverters enable the integration of solar power plants into the power grid, as the electrical systems use alternating current. In large plants, multiple central inverters are typically used to support the production and distribution of electricity.

To enable energy transmission at a higher level, a transformer for increasing the voltage to the medium voltage is used. This transformer raises the inverter's output voltage to transmission line, the resulting overvoltage wave travels along the line and reaches the solar power plant′s connection point with the transmission grid. This phenomenon can significantly impact the operation of the plant, particularly its step-up transformer, which is a critical and vulnerable component. The illustration also reflects the core objective of this study: to analyze the effects of lightning- caused overvoltages on the solar power plant and to evaluate the effectiveness of protective measures. Two case studies were conducted to assess this impact—one without surge arresters and another with surge arresters installed. In all simulations, a lightning strike was applied directly to the phase conductor of the transmission line to systematically examine the system′s response under varying conditions. This illustration represents the main objective of this study: to examine how overvoltages caused by lightning discharges on transmission lines can influence the operation of the solar power plant and to propose protective measures to mitigate these effects.

Regions with a high number of sunny days often coincide with areas experiencing significant thunderstorm activity. For example, the Croatian island of Brač enjoys approximately 2,600 hours of sunshine annually but also records around 40 lightning days per year [7, 8]. Similarly, the city of Trebinje in Bosnia and Herzegovina also boasts 2,600 annual sunshine hours while exhibiting very high lightning activity. Particularly noteworthy is the region of Lovćen in Montenegro, which experiences 50 lightning days annually alongside approximately 2,300 hours of sunshine [8, 9].

Given these conditions, it is highly justified to analyze the impact of lightning discharges on solar power plants in such regions, as understanding these effects is crucial for ensuring the reliability and safety of solar energy systems in areas prone to both high solar potential and frequent lightning activity.

### III. EMTP Modelling for Analyzing the Impact of Lightning-Caused Grid Disturbances on Solar Power Plants

In this section, the EMTP model developed for analyzing the impact of lightning-caused disturbances on solar power plants connected to the transmission grid is presented. The model includes a 110 kV transmission line, to which a solar power plant is connected. A detailed representation of the transmission line has been implemented, incorporating phase conductors, shield wires, towers, insulators, surge arresters and tower footing resistance. The remainder of the network has been modeled using a Thévenin equivalent.

The solar power plant is represented with a detailed EMTP model to accurately capture its interaction with the transmission line during lightning-caused overvoltage events.

The CIGRE model for lightning current has been employed to simulate lightning strokes.

Each of these components, along with their modeling approaches, is described in detail in this section.

#### A. EMTP model of the solar power plant

The EMTP model of the photovoltaic (PV) system includes several key components: PV panels, low-voltage (LV) to medium-voltage (MV) transformers, an equivalent PI circuit to represent the collector network, and MV to high-voltage (HV) transformers. A central element of the system is the photovoltaic module, often referred to as a panel, which consists of cells connected in series. This series configuration is essential for achieving the required voltage, current, and power output of the PV system. The model represents a 70 MVA solar power plant connected to a 110 kV transmission network. This integration ensures that the simulation accurately reflects both the power generation capacity of the photovoltaic system and its connection to the higher voltage transmission network, allowing for realistic analysis of the system's behavior in various operating conditions.

The detailed design of the model ensures an accurate and thorough representation of the PV system's performance within the EMTP simulation framework. The model replicates real-world conditions by integrating various elements that play specific roles in the system's operation.

The solar park transformer adjusts the output voltage from the PV panels to match the grid voltage or the inverter's requirements. Additionally, the inverter transformer modifies the inverter's output voltage to ensure compatibility with the grid before energy is delivered.

Each component of the model is crucial for its precision and reliability, enabling the simulation to account for diverse operating conditions and scenarios that the PV system might encounter. In table 1, used photovoltaic power plant transformer data is presented, while table 2 contains used inverter transformer data.

TABLE I. USED PHOTOVOLTAIC POWER PLANT TRANSFORMER DATA

| Connection type | Yd5 |
|---|---|
| Rated Power (MVA) | 70 |
| Rated Frequency (Hz) | 50 |
| Line-line voltage at the inverter side (kV) | 20 |
| Line-line voltage at the grid side (kV) | 110 |
| Resistance R (p.u.) | 0 |
| Reactance X (p.u.) | 0.11 |
| Impedance ratio of the windings 1 and 2 | 0.968 |

TABLE II. USED INVERTER TRANSFORMER DATA

| Connection type | Yd5 |
|---|---|
| Rated Power (MVA) | 3.5 |
| Rated Frequency (Hz) | 50 |
| Line-line voltage at the converter side (kV) | 0.575 |
| Line-line voltage at the grid side (kV) | 20 |
| Resistance R (p.u.) | 0 |
| Reactance X (p.u.) | 0.06 |
| Impedance ratio of the windings 1 and 2 | 0.998 |

#### B. Transmission line model

The transmission line in the model is represented by several essential components, including towers, insulators, phase conductors, shield wires, line surge arresters, and the tower footing resistance.

To accurately simulate the shielded transmission line, a multiphase line model was employed. This model incorporates several critical parameters to ensure precise calculations during transient simulations. These parameters include the physical dimensions of the phase conductors and shield wires, their DC resistance, and the (x, y) coordinates of each conductor and shield wire. Additionally, the vertical height of the conductors at the midspan above the ground, the ground return resistivity, and the overall line length are considered.

Table III provides a detailed overview of the electrical and geometrical properties of the conductors for simulated power line. The transmission lines were modeled using the EMTP CP line (Constant Parameter Line) element along with the transmission line data block.

TABLE III. DATA FOR OVERHEAD LINE

|  | DC resistance (Ω/km) | Outside diameter (cm) | x (m) | y (m) | y – with sag (m) |
|---|---|---|---|---|---|
| Phase A | 0.1444 | 1.708 | 2.5 | 22.7 | 14.1 |
| Phase B | 0.1444 | 1.708 | -3 | 20.5 | 11.9 |
| Phase C | 0.1444 | 1.708 | 3.5 | 18.3 | 9.7 |

For modeling the tower footing resistance, a constant resistance value was used. The models of towers, insulators, and surge arresters will be further explained in more detail in the continuation of this subsection to provide a comprehensive understanding of their roles and behavior within the transmission line system.

Although the lightning response of a transmission line tower involves electromagnetic phenomena, towers are typically modelled using circuit-based representations. This approach involves constructing the tower model with line sections and circuit elements that reflect its structural design. Such a method is widely adopted because it can be seamlessly integrated into general-purpose simulation tools like EMTP and is straightforward for practical engineers to interpret.

In this representation, the tower is divided into four main sections [10,11]:

- **Bottom Section**: This portion extends from the bottom crossarm to the ground and is modelled as a propagation element. It is characterized by its surge impedance and propagation length.
- **Top Section**: The section between the tower top and the top crossarm.
- **Intermediate Sections**: These are the segments between the crossarms.

The sections at the tower top (i.e., between the tower top and the top crossarm, as well as between the crossarms) are represented as inductive branches. This modelling approach enables the calculation of transient voltages at the tower top.

The tower's surge impedance is determined based on its geometric shape, and crossarms are included in the model with an inductance of 1 µH/m. Branch inductances for all sections are calculated based on the section length, tower surge impedance, and propagation velocity [12,13].

In the simulation model, insulators are represented as air gap elements, functioning based on the equal-area flashover model. A flashover occurs when the integral of the voltage across the insulator string reaches or exceeds a critical threshold D [10]:

$$\int_{t_0}^{t} (v_{gap}(t) - V_0)^K dt \geq D \quad (1)$$

where:

| | |
|---|---|
| $v_{gap}(t)$ | - voltage applied at the time t, to the terminals of the air gap; |
| $V_0$ | – minimum voltage to be exceeded before any breakdown process can start or continue; |
| K, D, V0 | - constants corresponding to an air gap configuration and overvoltage polarity |
| t (µs) | – time to flashover; |

For surge protection, a ZnO arrester model is employed. This model is based on the arrester's current-voltage (U-I) characteristics provided by the manufacturer. The following parameters were used for the simulated arrester:

- Rated voltage: 108 kVrms
- Maximum Continuous Operating Voltage (MCOV): 86 kV
- Nominal discharge current: 10 kA
- IEC class: II

The non-linear behavior of the arrester is represented by its U-I characteristic, ensuring an accurate simulation of its response under surge conditions. The EMTP tool incorporates these characteristics and models the arrester in the time domain as a nonlinear element. Its behavior is solved iteratively until convergence is achieved, based on the relative tolerance specified in the simulation parameters.

Additionally, the grounding rope is modeled with an inductance of 1 µH/m to simulate the transient response accurately.

*C. Network model*

The network model includes a three-phase source with a maximum line voltage of 123 kV. The model also incorporates positive- and zero-sequence impedances. The ratio of reactance to resistance of the analyzed power system is defined by the coefficient $\alpha$ calculated using Equation (2) [14,15]:

$$\alpha = \frac{X}{R} \quad (2)$$

The positive-sequence impedance ($Z_d$) and zero-sequence impedance ($Z_0$) are determined using the following equations [14,16]:

$$Z_d = \frac{U_{max}^2}{S_{tpsc}} = \frac{U_{max}}{\sqrt{3} I_{tpsc}} \quad (3)$$

$$Z_0 = U_{max}^2 \left( \frac{3}{S_{spsc}} - \frac{2}{S_{tpsc}} \right) = \frac{U_{max}}{\sqrt{3}} \left( \frac{3}{I_{spsc}} - \frac{2}{I_{tpsc}} \right) \quad (4)$$

Here, $U_{max}$ represents the maximum system voltage, $S_{tpsc}$ and $I_{tpsc}$ are the three-phase short-circuit power and current, while $S_{spcs}$ and $I_{spcs}$ present the single-phase short-circuit power and current, respectively.

Using these sequence impedances, the corresponding resistance and reactance values for both positive and zero sequences can be computed as follows:

$$R_d = \frac{Z_d}{\sqrt{1 + \alpha^2}} \quad (5)$$

$$X_d = \alpha R_d \quad (6)$$

$$R_0 = \frac{Z_0}{\sqrt{1 + \alpha^2}} \quad (7)$$

$$X_0 = \alpha R_0 \quad (8)$$

*D. Lightning stroke model*

The lightning stroke model is represented using the CIGRE lightning current model, which is integrated into the EMTP software. This model is widely recognized for its accuracy in representing the characteristics of lightning currents. Key lightning current parameters include: peak current, front time, tail time, and steepness. The CIGRE lightning current model, was utilized to simulate the impact of these parameters on the system under analysis. Further details about the CIGRE lightning current shape and its implementation can be found in [13].

## IV. SIMULATION SCENARIOS AND RESULT ANALYSIS

In this section, various simulation scenarios are presented to evaluate the impact of lightning-caused disturbances on the operation of the solar power plant.

If a lightning stroke hits the transmission line to which the solar power plant is connected, the resulting overvoltage wave propagates along the line and can disrupt the operation of the plant. Of particular concern is the step-up transformer, a critical and costly component of the system, which is especially vulnerable to these disturbances.

To evaluate the impact of lightning discharges on the solar power plant, two case studies were conducted: one without surge arresters and the other with surge arresters installed. For each case study, a series of simulation sets was performed to systematically analyse the system's response to lightning strikes under varying conditions. In all simulations, the lightning strike was applied directly to the phase conductor of the transmission line. The lightning current parameters used in the simulation sets 1-3 were selected based on values commonly reported in the literature [11], [13].

**Simulation Set 1:**
- **Lightning parameters:** Peak current of 31 kA, front time of 3 µs, and tail time of 75 µs.
- **Lightning strike locations:** Simulated at distances of 100 m, 300 m, 500 m, 700 m, 900 m, 1100 m, 1300 m, and 1500 m along the transmission line.

**Simulation Set 2:**
- **Lightning parameters:** Peak current of 10 kA, front time of 3 µs, and tail time of 75 µs.
- **Lightning strike locations:** Same as in Simulation Set 1.

**Simulation Set 3:**
- **Lightning parameters:** Peak current of 50.4 kA, front time of 3 µs, and tail time of 75 µs.
- **Lightning strike locations:** Same as in Simulation Set 1.

To accurately determine the parameters required for Simulation Set 3, a comprehensive analysis of the lightning performance of the transmission line was conducted using the Sigma SLP software developed by Sadovic Consultant [17]. This process involved calculating optimal lightning current parameters and evaluating the transmission line's response to lightning.

Electro-geometric modeling was employed to identify the strike locations along the transmission line. A total of 20,000 lightning strokes were analyzed, and the results were scaled to correspond to a 100 km length of the transmission line. Key findings from this analysis are presented in Table IV, where the maximum shielding failure current ($I_{max}$) was determined to be 50.4 kA, a critical parameter for our simulations.

To ensure accurate modeling of the line's lightning performance, an extensive number of electromagnetic transient simulations were carried out. Specifically, 3,000 simulations were performed, yielding detailed insights into the behavior of the transmission line under lightning-caused disturbances. The results, also converted to a 100 km line length for consistency, are summarized in Table V.

TABLE IV. CALCULATED PARAMETERS BASED ON ELECTRO-GEOMETRIC MODELLING

| Parameter | Value | Unit |
|---|---|---|
| Line attractive width | $W_E$ = 180.62 | m |
| Number of strokes collected by the line | $N_L$ = 45.6 | Strokes/100 km /year |
| Median current | $I_M$ = 29.8 | kA |
| Maximum shielding failure current | $I_{MAX}$ = 50.4 | kA |

TABLE V. EVALUATION OF LINE LIGHTNING PERFORMANCE: [STROKES/100 KM/YEAR]

| SFR | SFFR | BFR |
|---|---|---|
| 1.98 | 1.98 | 32.28 |

where is: SFR - Shielding Failure Rate, SFFR - Shielding Failure Flashover Rate and BFR - Back flashover rate.

Figure 2 shows examples of overvoltage caused by lightning strokes to the phase conductor, both in cases where a surge arrester is not installed and when a surge arrester is present. The presented case considers a lightning current of 50.4 kA, identified as the maximum shielding failure current based on numerous simulations conducted using the electro-geometric model. The waveform parameters for the lightning current are a front time of 3 µs and a tail time of 75 µs. In this scenario, the lightning stoke occurred at a distance of 100 meters from the solar power plant.

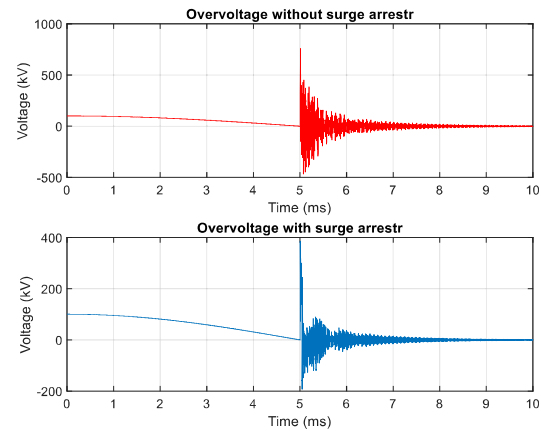

Fig.2. Impact of lightning stroke on phase conductor: with and without surge arrester

Maximum overvoltages provide a basis for comparing the two scenarios—one with a surge arrester installed and the other without—which are illustrated in Figures 3 and 4. The presence of a surge arrester significantly mitigates overvoltages, highlighting its crucial role in protecting the system. The results also show that overvoltages decrease as the distance between the lightning strike location and the solar power plant increases. Conversely, overvoltages rise with an increase in the lightning peak current. The most critical scenario occurs with a lightning peak current of 50.4 kA, where the absence of a surge arrester leads to particularly severe overvoltages, underscoring the necessity of installing such protective devices to enhance the resilience of solar power plants.

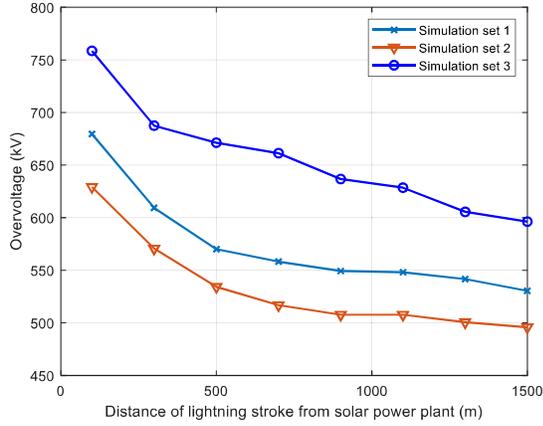

Fig.3. Overvoltages for different strokes and positions along the transmission line – case without surge arrester

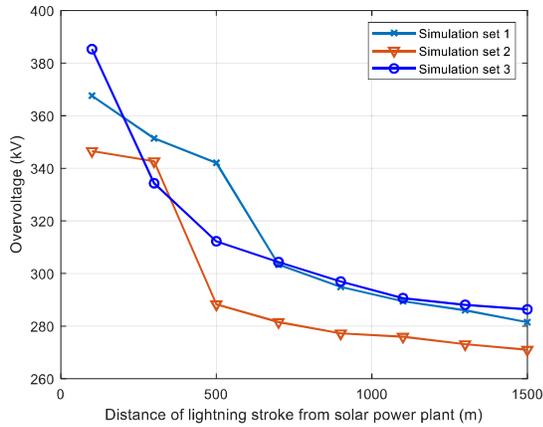

Fig.4. Overvoltages for different strokes and positions along the transmission line – case with surge arrester

Table VI presents the surge arrester currents for various lightning currents and distances from the solar power plant. The results indicate that the observed values are within acceptable limits, confirming that the selected surge arrester class, with a nominal discharge current of 10 kA, meets these requirements. However, it is important to note that simulations reveal that for a lightning current of 100 kA, the surge arrester current exceeds 60 kA, which could result in fact that the arrester blow up. Additionally, lightning strokes with a very steep front time of 1.2 μs can cause overvoltages reaching several megavolts, posing significant risks to system integrity.

TABLE VI. COMPUTED CURRENTS OF SURGE ARRESTER

| Distance (m) | $I_{ar}$ (kA) Simulation set 1 | $I_{ar}$ (kA) Simulation set 2 | $I_{ar}$ (kA) Simulation set 3 |
|---|---|---|---|
| 100 | 1.3211 | 1.2227 | 2.8556 |
| 300 | 1.0860 | 1.1172 | 1.9708 |
| 500 | 1.0572 | 1.0300 | 1.7700 |
| 700 | 1.0420 | 1.0260 | 1.6919 |
| 900 | 1.0364 | 1.0100 | 1.3852 |
| 1100 | 1.0281 | 1.005 | 1.2965 |
| 1300 | 1.0143 | 0.9850 | 1.1313 |
| 1500 | 1.0106 | 0.9810 | 1.0235 |

Given that the maximum shielding failure current was determined to be 50.4 kA, simulations were conducted using various front times, as shown in Table VII, while keeping the tail time constant at 75 μs. Overvoltages were calculated for these conditions. Subsequently, overvoltages were evaluated for a front time of 3 μs, with the tail time varying as presented in Table VIII. It can be concluded that an increase in the front time results in a decrease in overvoltage, whereas an increase in the tail time leads to higher overvoltages.

TABLE VII. COMPUTED OVERVOLTAGES FOR VARIOUS FTONT TIMES

| Front time $t_f$ (μs) | *Overvoltage* (kV) |
|---|---|
| 2 | 770.2911 |
| 3 | 758.6001 |
| 5 | 672.0400 |
| 8 | 596.6477 |

TABLE VIII. COMPUTED OVERVOLTAGES FOR VARIOUS TAIL TIMES

| Tail time $t_f$ (μs) | *Overvoltage* (kV) |
|---|---|
| 75 | 758.6001 |
| 150 | 765.3144 |
| 300 | 769.6020 |
| 500 | 771.2932 |

After determining that the maximum shielding failure amplitude of the lightning current is 50.4 kA, a Fourier transform of the resulting overvoltage signal caused by this lightning strike was performed. The entire analysis was conducted based on a single realization. The Fourier transform of the signal is presented, focusing on the magnitude spectrum. The sampling frequency used for the analysis was 100 MHz. The amplitude spectra of the signal with and without surge arrester are compared side by side.

Figure 5 illustrates the results of this analysis. In the low-frequency range, there was no significant impact on the frequency components of the signal, as observed for frequencies up to 10 kHz. However, the value of the DC component of the signal was noticeably higher.

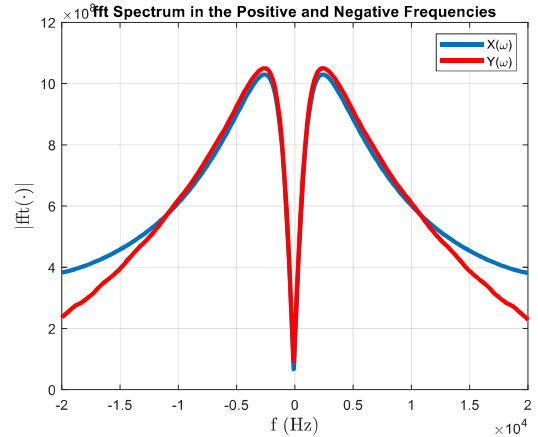

Fig.5. Frequency spectrum comparison of overvoltage signals without surge arrester X(ω) and with surge arrester Y(ω) in the low-frequency range

At frequencies above 10 kHz, an additional attenuation of the signal′s harmonic components can be observed, due to the presence of the surge arrester. The dominant frequency of approximately 55 kHz represents the oscillatory component present in the tail portion of the waveform, as depicted in Figure 6.

The surge arrester reduces the overall average signal power by approximately 5.33 times.

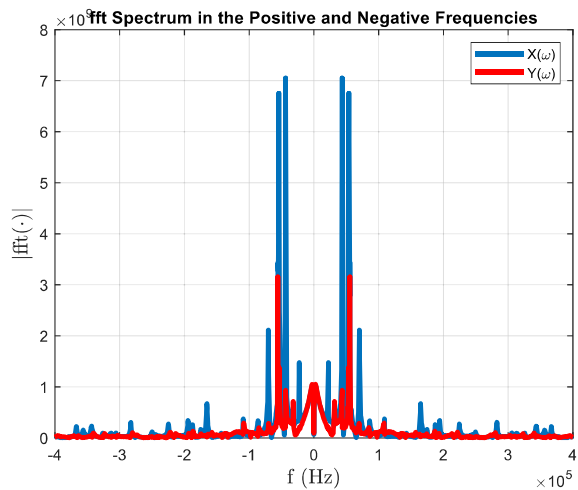

Fig.6. High-frequency spectrum analysis highlighting oscillatory components in overvoltage signals

Figure 7 illustrates the behavior of the surge arrester as a low-frequency filter with a 3-decibel passband width of approximately 18.2 kHz. Additionally, it demonstrates a noticeable amplification of the DC component's power by about two times (3 dB), aligning with the observations discussed earlier.

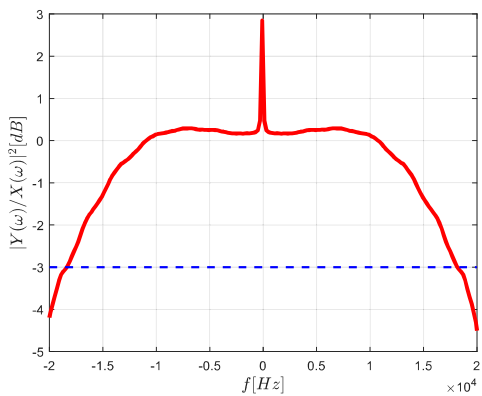

Fig.7. Surge arrester as a low-frequency filter

The Hilbert marginal spectrum is a significant tool in time-frequency analysis, particularly for characterizing the energy distribution of signals over frequencies. It is derived from the Hilbert-Huang Transform (HHT), combining Empirical Mode Decomposition (EMD) and Hilbert Spectral Analysis (HSA). While the EMD decomposes a signal into Intrinsic Mode Functions (IMFs), the HSA provides the instantaneous frequencies and amplitudes of these IMFs. The Hilbert marginal spectrum, as described in [18], represents the cumulative amplitude contribution of each frequency over the entire signal duration. This spectrum is particularly advantageous for analyzing non-stationary and nonlinear signals, which makes it a valuable tool for power system studies and transient analysis [19].

Figure 8 presents the Hilbert marginal spectrum for two voltage time series, one obtained without a surge arrester (blue line) and the other with a surge arrester (orange line).

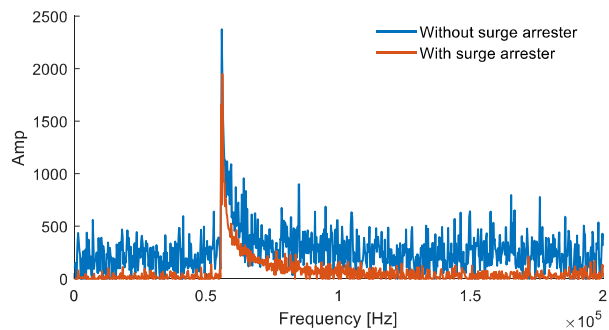

Fig.8. Hilbert marginal spectrum for two voltage time series, one obtained without a surge arrester and the other with a surge arrester

The x-axis represents the frequency content of the signals, showing the distribution of energy or amplitude contributions across different frequencies. The y-axis indicates the magnitude of the signal's energy contribution at each frequency, measured in amplitude (Amp).

The signal without the surge arrester exhibits significantly higher amplitudes across the frequency spectrum. This behavior reflects the absence of any damping or energy suppression mechanisms. The maximum amplitude for this signal is observed at a frequency of 55.863 kHz, where it reaches a peak value of 2375.85. This peak indicates a prominent resonance or high-energy contribution at this specific frequency, which is likely a result of the lack of protective measures against transient surges.

In contrast, the signal with the surge arrester shows considerably reduced amplitudes throughout the frequency spectrum. This demonstrates the effectiveness of the surge arrester in mitigating transient energy. Although the maximum amplitude for this signal is also observed at the same frequency of 55.863 kHz, the peak value is significantly reduced to 1950.89. This reduction in amplitude illustrates that the surge arrester effectively dampens the transient energy at this critical frequency.

The surge arrester plays a crucial role in protecting the system by significantly attenuating the harmonic components of the overvoltage and reducing the amplitude. This helps to prevent potential damage to the equipment and ensures the stability and reliability of the power system.

## V. Conclusions

This study provides a detailed analysis of the risks posed by lightning discharges to solar power plants connected to the transmission grid. Simulations of lightning strokes at various distances from the solar power plant evaluated the impact of overvoltages on system performance, considering scenarios both with and without surge arresters. Fourier analysis of overvoltages and the Hilbert marginal spectrum of voltage time series comparing cases with and without surge arresters highlighted the critical role of surge arresters in mitigating overvoltage effects. The findings demonstrate that surge arresters significantly reduce the impact of lightning strikes, protecting equipment and ensuring system reliability, particularly in regions prone to high lightning activity.

Based on these findings, the authors strongly recommend the installation of surge arresters as an essential protective measure for solar power plants connected to transmission lines. The results confirm that surge arresters effectively limit overvoltages and safeguard critical components such as step-up transformers, ultimately enhancing the overall resilience of the system. This underscores the necessity of integrating effective lightning protection measures into solar power plant designs. Future work should focus on advanced protection mechanisms and improved modeling techniques to further enhance the resilience and stability of solar power systems in lightning-prone areas.